\documentclass[5p]{elsarticle}
\usepackage{graphicx}
\usepackage{amssymb}
\usepackage{color}
\definecolor{blue}{rgb}{0,0,0}

\begin{document}

\title{Huge field-effect surface charge injection and conductance modulation in metallic thin films by electrochemical gating}

\author [poli] {M.Tortello}
\author [poli]{A. Sola}
\author [poli]{Kanudha Sharda}
\author [poli]{F. Paolucci\fnref{fn1}}
\author [poli]{J.R. Nair}
\author [poli]{C. Gerbaldi}
\author [poli]{D. Daghero}
\author [poli]{R.S. Gonnelli\corref{cor1}}
\ead{renato.gonnelli@polito.it}
\cortext[cor1]{Corresponding author}
\fntext[fn1]{Now at Max Planck Institute for Solid State
Research, Stuttgart (Germany).}
\address[poli] {Dipartimento di Scienza Applicata e Tecnologia, Politecnico di Torino, Corso Duca degli Abruzzi 24, 10129 Torino, Italy}

\begin{abstract}

The field-effect technique, popular thanks to its application in
common field-effect transistors, is here applied to metallic thin
films by using as a dielectric a novel polymer electrolyte solution.
The maximum injected surface charge, determined by a suitable
modification of a classic method of electrochemistry called
double-step chronocoulometry, reached some units in 10$^{15}$
charges/cm$^{2}$. At room temperature, relative variations of
resistance up to 8\%, 1.9\% and 1.6\% were observed in the case of
gold, silver and copper, respectively and, if the films are thick
enough ($\geqslant 25$ nm), results can be nicely explained within a
free-electron model with parallel resistive channels. The huge
charge injections achieved make this particular field-effect
technique very promising for a vast variety of materials such as
unconventional superconductors, graphene and 2D-like materials.

\end{abstract}

\maketitle


\section{Introduction}

One of the milestones of the technological revolution that  has
occurred in the last $50$ years is the invention of the field effect
transistor (FET). As it is well known, the working principle of FET
devices is based on the modulation of the transport properties of a
material by means of an applied electric field. This is important
not only from a technological point of view but also from a
fundamental one. Indeed, field-effect experiments allow tuning the
charge density of a material without the side effects typical of
chemical substitutions or application of pressure, such as
introduction of disorder or modification of the lattice structure.
In this regard field-effect experiments allowed, for instance,
enhancing the critical temperature of some superconductors
\cite{Glover60,Mannhart91,Ahn99}, inducing metallic behavior in
insulators \cite{Shimotani07}, {\color{blue} {a metal-to-insulator transition in a colossal magnetoresistive manganite \cite{Dhoot09}}} or even a superconducting phase transition in materials like SrTiO$_{3}$ \cite{Ueno08}, ZrNCl \cite{Ye09}, and KTaO$_{3}$ \cite{Ueno11}.
In order to achieve large effects, a high amount of charge has to be injected and therefore very intense electric fields are required. Instead of using the standard solid dielectric which is nowadays a very common solution for commercial devices, the use of a polymer electrolyte solution (PES) \cite{Shimotani07,Dhoot09,Ueno08,Ye09,Ueno11,Panzer05,Dhoot06} is a very promising technique to achieve these huge charge injections. With a solid-dielectric device it is possible to induce a surface charge $n_{2D}^{max}$ of the order of $10^{13}$ charges/cm$^{2}$ while with polymeric gating techniques this value can easily reach some units in $10^{14}$ charges/cm$^{2}$ as a consequence of electric fields as high as $100$ MV/cm. {\color{blue}{By using a suitable PES, Dhoot and coworkers have recently achieved the unprecedented surface density of injected charge of $2 \times 10^{15}$ charges/cm$^2$ \cite{Dhoot09}.}}
The electrochemical gating technique is based on the formation of an electric double layer (EDL) between an electrolyte solution and the surface of the sample under test: the polymer solvates positive and negative ions in the electrolyte solution and an external bias drives them towards the oppositely polarized surface of the film or gate, thus forming the EDL. Therefore, the EDL acts as a parallel-plate capacitor with extremely small distance between the plates (of the order of the polymer molecule size) \cite{Ye09} and thus a very large capacitance.\\
As already stated above, field-effect experiments have been performed in some exotic materials in order to induce dramatic modifications of their properties or even phase transitions \cite{Ahn99}. However, although poorly studied, field-effect measurements have been carried out also in metals. The little interest in this topic is mainly due to the fact that it seems not to be so much attractive from an applicative point of view, but also because the effect is commonly considered to be difficult, or almost impossible, to observe: indeed, the electronic screening length in the semiclassical model for a metal is less than one atomic radius. Nevertheless, field-effect induced modulations of the conductivity in metals have already been observed in the past \cite{Glover60,Bonfiglioli59,Bonfiglioli65,Stadler65,Martinez97,Markovic01,Daghero12}, also justifying a fundamental interest in this subject.\\
Here we report on field-effect experiments performed on gold, silver and copper by means of the electrochemical gating technique. The novel PES we adopted allows surface charge injections as high as {\color{blue}{$4.5 \cdot 10^{15}$ charges/ cm$^{2}$ \cite{Daghero12}(thus even larger than those obtained by Dhoot et al. \cite{Dhoot09})}} and far-from-negligible modulations of the conductivity were observed in all the investigated metals. In particular, relative variations of the resistivity $\Delta R / R^\prime$ up to 8 \%, 1.9\% and 1.6 \% were achieved at room temperature in gold, silver and copper, respectively. If films are not too thin, such as to avoid a dominant role of surface scattering, the trend of $\Delta R / R^ \prime t$ as a function of $n_{2D}$ (where $t$ is the thickness of the film) can be nicely explained within the free-electron model. Moreover, the huge surface charge injections achieved make this technique promising for many other materials such as graphene, unconventional superconductors and 2D-like crystals.

\section{Experimental setup}

\begin{figure}[t]
\begin{center}
\includegraphics[keepaspectratio, width=0.8\columnwidth]{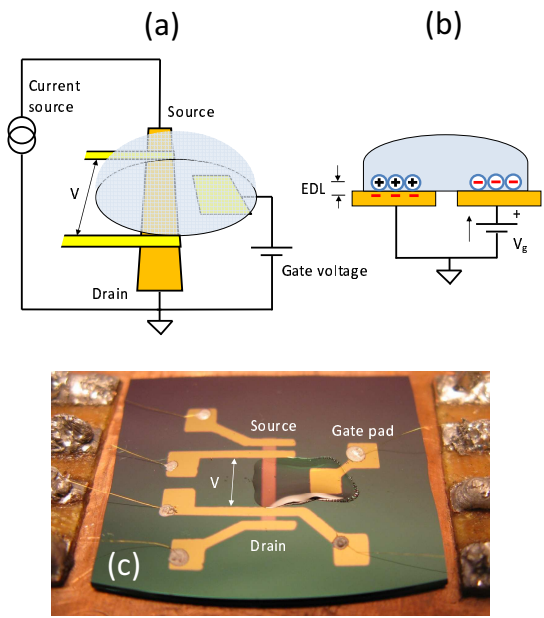}
\end{center}
\vspace{-5mm}\caption{a) Scheme of devices for electrochemical gating: the portion of the film between the voltage contacts is covered by the PES that is polarized by a voltage source (V$_{\textrm{gate}}$). The film resistivity is measured in the standard four-probe configuration. (b) Cross-section sketch of the device: when V$_{\textrm{gate}}$ is applied, ions are accumulated at the interface with the metallic film and a charge with opposite sign is induced on its surface, forming an EDL. (c) Photograph of a typical device used in our field-effect experiments on metallic thin films.} \label{fig:1}
\end{figure}

\begin{figure}
\begin{center}
\includegraphics[keepaspectratio, width=\columnwidth]{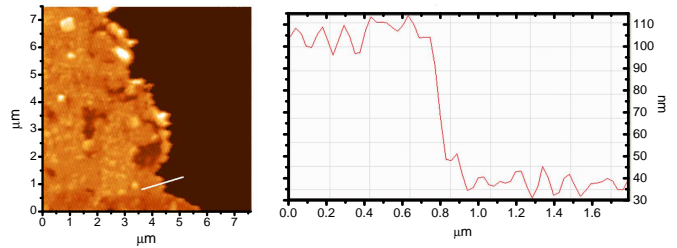}
\end{center}
\caption{{\color{blue}{Left: an example of AFM image of the surface of a Ag film. The flat right-hand side of the image is the glass substrate. The sharp edge shown here was obtained by scratching the film. The white straight line indicates the cut along with the z-profile shown in the right panel was measured. This profile indicates a thickness of $65 \pm7$ nm. The effective thickness of this film that results from the correction for voids as described in ref.\cite{Rowell03} and in the text is equal to 30 nm.}}}\label{Fig:AFM}
\end{figure}

Figure \ref{fig:1}(a) shows a sketch of the experimental setup. The device, fabricated in the planar configuration as in ref.\cite{Dhoot06}, consists of the metallic thin film under study, the gate pad and the PES. Thin films are obtained by physical vapor deposition (PVD) of the selected metal (gold, silver or copper) at a pressure $P \sim 2 \cdot10^{-5}$ mbar. {\color{blue}{Their thickness is measured by means of atomic force microscopy (AFM) which is also used to investigate the morphology of the film surface, together with field-emission scanning electron microscopy (FESEM)\cite{Daghero12}. To account for the possible presence of voids between grains that can reduce the effective cross sectional area of the film, we also applied a correction proposed by Rowell \cite{Rowell03} and based on the fact that the temperature dependence of the resistivity depends only on the material and not on the form of the sample (bulk, polycrystal, film). The procedure requires comparing the experimental $\rho(T)$ of a given film with the resistivity of the pure (bulk) material, and  provides a scaling factor $F$ for the geometrical cross section. In most Au films, we found out that $F\simeq 1$, i.e. there is no need of correction and the geometrical cross section coincides with the effective one. In all the Cu and Ag films, instead, the correction was necessary, which is consistent with AFM and FESEM images that show a greater roughness. An example of AFM image on a Ag film and the corresponding z-profile to determine its thickness is shown in Figure \ref{Fig:AFM}.}}

Voltage, current and gate pads were deposited on top of the film by evaporating gold, independently of the metal under study. As a substrate, we adopted {\color{blue}{either glass, or amorphous SiO$_{2}$ or Si$_{3}$N$_{4}$ on a Si wafer.}}

The metallic thin film behaves as the channel of a standard FET whose resistivity is measured with a standard four-probe configuration by inverting the current during each measurement or by using the ac technique. The PES acts as the dielectric and is therefore connected on one side to the metallic film and on the other side to the gate pad. A bias is applied to the gate by means of a source-measure unit that, at the same time, measures the electric current flowing through the PES. The PES we used was obtained by a reactive mixture of bisphenol A ethoxylate (15 EO/phenol) dimethacrylate (BEMA; average Mn: 1700, Aldrich), poly(ethylene glycol)methyl ether methacrylate (PEGMA; average Mn: 475, Aldrich), and lithium bis(trifluoromethanesulfonyl)imide (LiTFSI) in the presence of $3\%$ wt of a 2-hydroxy-2-methyl-1-phenyl-1-propanon free radical photoinitiator (Darocur 1173, Ciba Specialty Chemicals). The quantities of BEMA and PEGMA are in a 3:7 ratio, and the LiTFSI is the 10$\%$ wt of the total compound. The PES was then polymerized by UV exposure using a medium vapor pressure Hg UV lamp, with a radiation intensity on the surface of the sample of 30 mW/cm$^{2}$. All the above operations were performed in the controlled Ar atmosphere of a dry glove box with O$_{2}$ and H$_{2}$O content $<0.1$ ppm. {\color{blue}{Since the PES used here was originally developed for Li-ion battery applications, the reproducibility of its chemical and physical properties was widely tested by means of conductivity measurements, crosslink density measurements, thermal history and cyclic voltammetry \cite{Nair11}. A careful preparation according to the above recipe results in a very high reproducibility and stability of the PES over long times (several months).}}

{\color{blue}{Figure \ref{fig:1}(b) shows how the field-effect experiments work}}: The application of the gate voltage causes an accumulation of charge at the interface between the film and the PES, thus forming the EDL. A symmetric amount of charge is induced at the gate pad, shown on the right hand-side of Fig. \ref{fig:1}(b). Figure \ref{fig:1} (c) shows a photo of a real device used in these experiments.

\section{Experiment and models}

In this section we describe the measurements performed on the devices previously shown, as well as the models used to determine the injected surface charge and to quantify the effect of surface doping on the conductivity of the metallic thin films.

\begin{figure}[t]
\vspace{-10mm}
\begin{center}
\includegraphics[keepaspectratio, width=0.8\columnwidth]{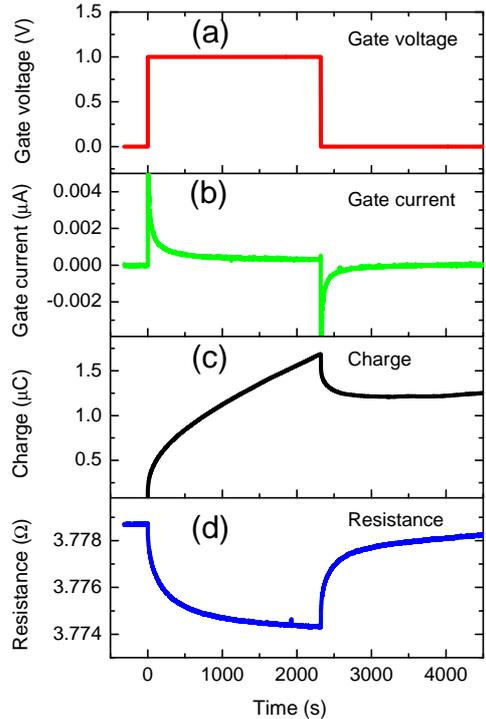}
\end{center}
\vspace{-7mm}\caption{a) Gate voltage vs time.
b) Current flowing through the PES during the application and removal of gate voltage. c) Charge moved during the measurement, obtained by integration of the current shown in b). d) Effect of application of the gate voltage on the thin film resistance.} \label{fig:2}
\end{figure}

The sequence of operations and measurements performed on a single device is summarized in Figure \ref{fig:2}.  At $t=0$ a voltage is applied between the gate pad and the film. After a proper time (of the order of several hundreds of seconds due to the rather slow dynamic of the PES) this voltage is removed as shown in Figure \ref{fig:2}(a). A sharp peak in the gate current flowing through the PES appears immediately after the application of gate voltage, indicating the formation of the EDL. This current decreases almost exponentially and after a few seconds it reaches a very small constant value. When the gate voltage is removed a similar peak appears with negative values of the current, indicating the disruption of the EDL (Figure \ref{fig:2}(b)). The time integral of this gate current as function of time (Figure \ref{fig:2}(c)) provides {\color{blue}{the total amount of charge that flows through the PES and, by applying a proper procedure, allows extracting the (smaller) charge that is involved in the formation of the EDL}}. Finally, this dynamical formation and destruction of the EDL with the consequent charge injection at the surface of the metallic thin film {\color{blue}{gives rise to}} relative changes in the film resistance of the order of 10$^{-3}$ as shown in Figure \ref{fig:2}(d). As expected, the application of a positive bias to the gate induces a negative charge at the film surface (see Fig. \ref{fig:1}(b)) giving rise to a reduction of the total resistance of the film. In order to explain the experimental results just described, it is first necessary to carefully determine the surface charge injected in the film by the polymer gating and then to calculate the effect of this charge on the transport properties of the whole film. Both these topics are addressed in the following.

In field-effect experiments performed on very thin films of materials less conductive than metals a direct estimation of charge injection can be done by means of Hall-effect measurement at the sample surface. In metallic thin films this is not possible as it would require huge magnetic fields because of the high intrinsic carrier density of the metal and the considerable thickness of the film (here of the order of {\color{blue}{some tens of nanometers}}). In the absence of a direct method for estimating the total charge on the side of the film we must rely on the determination of the additional charge induced on the film surface through the quantification of the EDL charge on the PES side. A rough integration of the current flowing through the PES would give an overestimation of the charge that forms the EDL. In fact, determining the charge of the EDL by integrating the gate current is not correct if electrochemical effects are present, as pointed out in Ref. \cite{Yuan10}. Indeed, in the current vs time graph it is possible to observe a slowly vanishing current that always persists long after the application of the gate voltage (see Fig. \ref{fig:2}(b)). This current is due to the flow of charges necessary to maintain the gradient of ion concentration when diffusion of electroreactants \cite{Inzelt10} or tunneling effects through the EDL \cite{Yuan10} take place. {\color{blue}{However, being several orders of magnitude smaller than the direct current flowing between source and drain, this gate current does not affect the resistance measurements}}.

\begin{figure}[t]
\begin{center}
\includegraphics[keepaspectratio, width=\columnwidth]{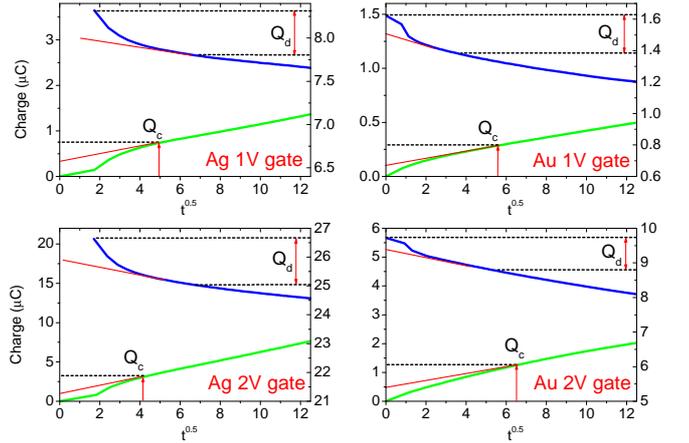}
\end{center}
\vspace{-5mm}\caption{Four different examples of chronocoulometry plots for two different materials and two different gate voltages. In each plot, thick lines represent the electric charge obtained by integrating the gate current in the charging (green bottom line) and discharging (blue top line) phases. Please note the different vertical scales (left for charging and right for discharging).} \label{fig:3}
\end{figure}

A suitable method of electrochemistry used to evaluate \emph{only} the charge that forms the EDL (which, in principle, corresponds to the charge injected in the metallic film), separating it from the charge related to other phenomena, is called double-step chronocoulometry \cite{Inzelt10}. The starting point is the time dependence of the calculated charge $Q(t)=\int_{0}^{t}I_{G}(t')dt'$ shown in Fig. \ref{fig:2}(c). This $Q(t)$ dependence can be analyzed both during the application and the removal of gate voltage, giving two complementary {\color{blue}{pieces of}} information about the charge in the EDL. The shape of $Q(t)$ shows indeed that two phenomena occur on very different time scales: a rather fast EDL charging/discharging (that is expected to show an exponential time dependence) and other effects of electrochemical nature that should give a $\sqrt{t}$ dependence \cite{Inzelt10}. By plotting $Q$ as function of $\sqrt{t}$ (see Fig. \ref{fig:3}) it is very easy to determine the instant $t^{*}$ at which the $\sqrt{t}$ behavior becomes dominant, i.e. when the exponential behavior has become negligible.  We assume that the total charge injected in the film is $Q(t^*)$. This definition is different from the one used in standard chronocoulometry, where the EDL charge is obtained by the intercept of the linear fit of the $Q$ vs $\sqrt{t}$ dependence with the vertical axis. The physical reason is the following: Standard chronocoulometry procedure implies that the starting of the current that generates the EDL and that of the current due to electrochemical phenomena are concurrent and that the scale of time in which these currents develop is the same and very small. In the case of our viscous PES the diffusion of electroreactants is likely to be rather slow and, consequently, it is reasonable to expect that the current due to electrochemical phenomena starts with a certain delay and reaches its maximum value sometimes later compared to that which forms the EDL. This is exactly the kind of approximation we adopted in assuming that $Q_{EDL}=Q(t^{*})$. This double-step procedure has the advantage to return both the charge that builds up the EDL during the application of gate voltage and the charge that comes from the dissolution of the EDL, during the removal of gate voltage, allowing a consistency check between the measured values ​​at two different moments.

In Figure \ref{fig:3} four double-step chronocoulometry plots are presented for two different metallic films (silver and gold) and for two different gate voltages. Each graph shows the charging (thick solid line, bottom) and the corresponding discharging (thick solid line, top) curve plotted with respect to the square root of time. The main features previously described are clearly visible: a rapid increase (decrease) of the charge is present in the first 15-40 seconds after the application (removal) of the gate voltage, corresponding mainly to the phase of formation (destruction) of the EDL. Subsequently the $Q$ vs $\sqrt{t}$ curves become linear with positive (negative) slope indicating that a phase dominated by the diffusion of electroreactants has been reached. The linear fits of the different curves are shown in Fig. \ref{fig:3} as red thin lines. The discharging curves (vertical scale on the right) are symmetrical with respect to the charging ones but, of course, start at high $Q$ values: in this case the EDL charge is evaluated as the difference between the initial $Q$ at $t=$ 0 and the one reached when the curve becomes linear. Finally the EDL charges obtained during the charging phase ($Q_c$ in the plots) and the discharging one ($Q_d$ in the plots) are averaged to obtain a single value $Q_{EDL}$ for every applied gate voltage. It turns out that $Q_c$ and $Q_d$ are very similar to each other as can be seen in Fig. \ref{fig:3} (please note the different vertical scales) and, of course, they increase at the increase of $V_{gate}$. Let's now suppose that $Q_{EDL}$, measured by this modified version of double-step chronocoulometry, is the same charge (with opposite sign) $Q_{i}$ induced at the surface of the metallic thin film. The induced surface charge density will thus be $n_{2D}= Q_i/eS$, where $S$ is the surface of the film covered with the PES (gated area) and $e$ is the electronic charge.

We must now focus on where the charge induced by the polymeric gating is distributed and how it affects the total resistivity of the film. Since the charge that we induce on the metallic film is only (in a first, rough approximation) two-dimensional, we can describe the thin film as a parallel of a perturbed and an unperturbed 3D region, where the three-dimensional charge density of the perturbed region follows a density profile $n_{3D}(z)$, where $z$ is the axis normal to the interface and the density decays on a length scale defined by a parameter $\xi$. Both $n_{3D}(z)$ and $\xi$ depend on the material but for metallic films and within a simplified semiclassical model, one can imagine that the whole injected charge is uniformly distributed in a surface layer of thickness $\simeq\xi$ so that $n_{3D}=n_{2D}/ \xi$. Different choices are possible for $\xi$: from the classical Thomas-Fermi screening length that in good metals is of the order of 0.5 {\AA} to more complex expressions that take into account quantum-mechanical screening phenomena. Often a simple rule of thumb is used that consists in assuming the thickness of one atomic layer for the region where the charge is perturbed by the field. However, for our purposes here, the exact calculation of this quantity is irrelevant. In fact it can be shown both by a simple classic approach and by a much more complex perturbative self-consistent quantum model based on the Lindhard-Hartree theory of the
electronic screening \cite{Omini} that $\xi$ does not enter the final expression for the relative variation of the total film resistance.

The simple free-electron calculation of this relative variation $\Delta R / R^\prime$, that is based on the parallel of the perturbed surface region and of the unperturbed bulk one and assumes a constant effective electron mass and relaxation time, gives

\begin{equation}
\Delta R/R'=\frac{R(V_g)-R_0}{R(V_g)}=-\frac{n_{2D}}{n t}.
\label{eq:DeltaR}
\end{equation}

where $V_g$ is the gate voltage, $n$ is the unperturbed 3D density of charge carriers, $R_0$ is the unperturbed resistance when $V_g=0$ and $t$ is the total thickness of the film.

\section{Results and discussion}

\begin{figure}[t]
\begin{center}
\includegraphics[keepaspectratio, width=\columnwidth]{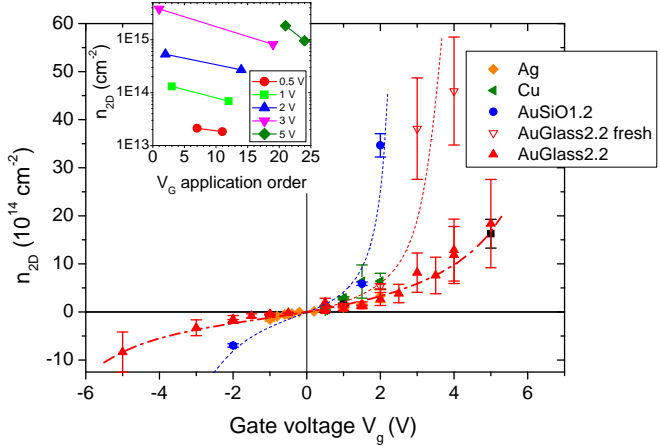}
\end{center}
\vspace{-5mm}\caption{Induced surface charge density $n_{2D}$ as function of the gate voltage $V_g$ for different devices made of different metals (Au, Ag, Cu). {\color{blue}{Although most of the points are gathered around a common trend, some data show much higher values of the surface charge density. Lines are only guides to the eye.}} In the inset the dependence of $n_{2D}$ on the sequence of application of $V_g$ is shown. For details see the text.} \label{fig:4}
\end{figure}

Figure \ref{fig:4} shows the dependence of the injected surface charge density $n_{2D}$ on $V_g$ determined by the modified double-step chronocoulometry in different materials and devices. For $|V_g|\lesssim 2$ V $n_{2D}$ depends almost linearly on $V_g$ and then it grows more than linearly for higher $V_g$. The injected charge densities for positive and negative gate voltages are rather symmetric for $|V_g|\lesssim 2$ V while for $|V_g|>2$ V the negative ones are smaller. It is interesting to notice that there are data of different metals (Au, Ag) and of different devices that are gathered around the same guide for the eyes {\color{blue}{(thick dash-dotted line) but, at the same time, there are some other data (relevant to different metals: Cu, Au) that for the same gate voltages indicate much higher $n_{2D}$ values (circles, open triangles, left triangles). Some of these data (open triangles) were obtained in the device called AuGlass2.2 that however also gave charge densities that lie on the lower trend line (solid triangles).}} Evidently the large differences in the $n_{2D}$ values measured at the same $V_g$ do not depend on the particular metal or on the specific device. To better understand the origin of this behaviour we analysed the sequence of $V_g$ applications in AuGlass2.2. {\color{blue}{The inset to Fig. \ref{fig:4} shows the values of $n_{2D}$ for different gate voltages as a function of the sequential number of the measurements. It is clear that, for any gate voltage, the charge injection is maximum in the first application and systematically decreases in the following. Indeed, the highest $n_{2D}$ values measured in AuGlass2.2 and shown in the main panel of Fig.\ref{fig:4} were obtained in the first few $V_g$ applications, i.e. in the ``fresh'' device.  The subsequent values of $n_{2D}$ lie on the main trend line common to other devices and materials (thick dash-dot line). Since the PES is, in itself, very stable over long times, this behaviour might be rather ascribed to a sort of ``memory'' effect -- probably related to the ``loss'' of Li ions at the interface with the electrodes -- that certainly limits the performances of the PES.}} Record $n_{2D}$ values of the order of 4.5$\cdot 10^{15}$ cm$^{-2}$ that at the beginning can be obtained with $V_g \sim 2-3$ Volt (see main panel of Fig. \ref{fig:4}) after some use of the device would be reached only by applying much higher $V_g$ with the consequent risks of possible reactions between the electrolyte and the sample.
{\color{blue}{Finally, the fact that the highest charge injections were obtained in Au films on SiO$_2$ but also on glass seems to exclude any effect of the substrate on the device performance, which looks rather reasonable. However, for the time being we do not have a sufficient statistics to properly separate the ``memory'' effect from that of the substrate.}}

\begin{figure}[t]
\begin{center}
\includegraphics[keepaspectratio, width=\columnwidth]{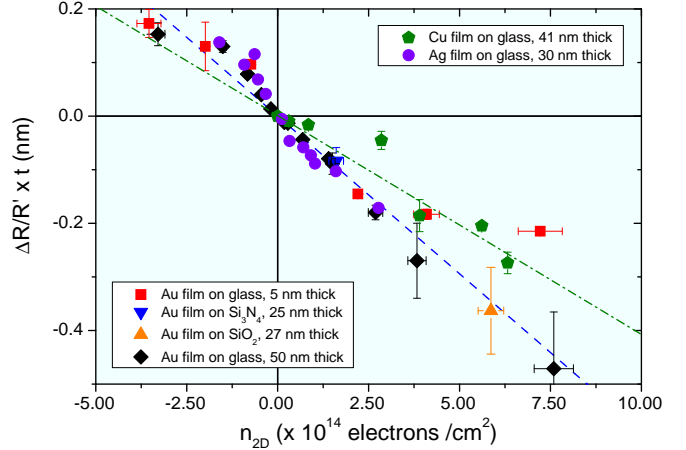}
\end{center}
\vspace{-5mm}\caption{Dependence of $\Delta R/R' \times t$ on $n_{2D}$ as obtained for films of different metals (Au, Ag, Cu) with different thickness and on different substrates (indicated in the legend). The straight lines are only guides to the eyes. } \label{fig:5}
\end{figure}

In Figure \ref{fig:5} the results concerning the film resistance variations generated by the charge injection are shown. According to eq. \ref{eq:DeltaR} the quantity $\Delta R/R' \times t$ should be a linear function of $n_{2D}$ with a negative slope. As a matter of fact Fig. \ref{fig:5} shows that this is the case for the variety of films of different metals and devices studied here. However the data for gold (full squares, triangles and diamonds) and silver (full circles) lie on the same straight blue dashed line which has a higher slope than the green dash-dotted line that approximates the data for copper (full pentagons).  The explanation appears quite simple: In eq. \ref{eq:DeltaR} the slope of the $\Delta R/R' \times t$ vs $n_{2D}$ curve is inversely proportional to the volume carrier density $n$ of the metal. But in the framework of the free-electron model the $n$ values of Au and Ag are quite similar (5.90 and 5.86 in units of 10$^{22}$ cm$^{-3}$, respectively) while $n$ for Cu is $8.47 \cdot 10^{22}$ cm$^{-3}$. The higher bulk carrier density of copper can nicely explain the reduced slope of the $\Delta R/R' \times t$ behaviour shown in Fig. \ref{fig:5}. Gold films with very low effective thickness (for example 5 nm, as in the case shown by red squares in Fig. \ref{fig:5}) deviate from the linear behavior at $n_{2D}$ greater than $2-3 \cdot 10^{14}$ cm$^{-2}$. This could be due to the increased role that scattering phenomena at the film surfaces have in ultrathin films. In these conditions a simple free-electron model that neglects the surface scattering may be no longer appropriate to represent the real physical situation. A reduction in the absolute value of $\Delta R/R'$ for a given $n_{2D}$ is indeed predicted by the already mentioned quantum perturbative model \cite{Omini}, when the probability of electron reflection at the surface is not negligible. The maximum absolute value of $\Delta R/R'$ obtained up to now in our polymer-gating field-effect experiments at room temperature is of the order of 8\%, 1.9\% and 1.6\% for Au, Ag and Cu, respectively. Low temperature experiments (down to 4.2 K) have shown that $\Delta R/R'$ values up to 10\% can be obtained in gold \cite{Daghero12}.

In conclusion, we have shown that {\color{blue}{in polymer-gating field-effect experiments it is possible to induce surface charge densities $n_{2D}$ up to $3-4 \cdot 10^{15}$ charges/cm$^{2}$ by using a suitable polymer-electrolyte solution. This value of $n_{2D}$ is approximately one order of magnitude greater than that obtained in almost all the similar experiments reported in literature, and even larger than the maximum achieved so far \cite{Dhoot09}}}. A proper modification of the classic double-step chronocoulometry method gives us a very versatile tool for the determination of the amount of surface charge induced in these experiments, even when the standard Hall-effect technique cannot be adopted as in the case of metallic thin films. The huge resistance shifts (up to 8\%) observed at room temperature in thin films of different metals (Au, Ag, Cu) reveal the power of this technique and, maybe, could suggest possible applications of these results. The linear dependence of the quantity $\Delta R/R' \times t$ on $n_{2D}$ observed in our thin films is explained very well by a simple free-electron model with parallel resistive channels, that also accounts for the observed differences in the slope of the above dependence in metals with a different {\color{blue}{intrinsic carrier density}}. The future application of this technique to conventional and unconventional superconductors, topological insulators and graphene \cite{Giustino10} or other 2D-like materials could lead to a large and defect-free modulation of their surface electronic properties with extraordinary consequences that now can only be imagined.\\

The authors acknowledge Francesco Laviano for AFM measurements on the films.

\section*{References}


\begin{thebibliography}{99}
%
\bibitem{Glover60} R. E. Glover, III and M. D. Sherrill, \textit{et al.} \textit{Phys. Rev.
Lett.} \textbf{5}, 248 (1960).
%
\bibitem{Mannhart91} J. Mannhart \textit{et al.}, \emph{Z. Phys.
B} \textbf{83}, 307 (1991).
%
\bibitem{Ahn99} C. H. Ahn \textit{et al.} \textit{Science} \textbf{284}, 1152
(1999).
%
\bibitem{Shimotani07} H. Shimotani \emph{et al.}, \textit{Appl. Phys. Lett.} \textbf{91}, 082106
(2007)
%
\bibitem{Dhoot09} A.S. Dhoot, C. Israel, X. Moya, N.D. Mathur, and R.H. Friend, \textit{Phys. Rev. Lett.} \textbf{102}, 136402 (2009).
%
\bibitem{Ueno08} K. Ueno et al., \textit{Nature Mater.} \textbf{7}, 855-858 (2008).
%
\bibitem{Ye09} J. T. Ye \textit{et al.} \textit{Nature Mater}. \textbf{9},
125-128 (2009).
%
\bibitem{Ueno11} K. Ueno et al., \textit{Nature Nanotech.} \textbf{6}, 408 (2011).
%
\bibitem{Panzer05} M. J. Panzer, C. R. Newman, and C. D. Frisbie,
\textit{Appl. Phys. Lett.} \textbf{86}, 103503 (2005)
%
\bibitem{Dhoot06} A. S. Dhoot \textit{et al}. \textit{Proc. Natl. Acad. Sci}. \textbf{32}, 11834-11837 (2006).
%
\bibitem {Bonfiglioli59} G. Bonfiglioli and R. Malvano, \textit{Phys. Rev.} \textbf{115}, 330 (1959).
%
\bibitem{Bonfiglioli65} G. Bonfiglioli, E. Coen and R. Malvano, \textit{Phys. Rev.} \textbf{101}, 1281 (1965).
%
\bibitem{Stadler65} H.L. Stadler, \textit{Phys. Rev. Lett.} \textbf{14}, 979 (1965)
%
\bibitem{Martinez97} G. Martinez-Arizala \textit{et al.}, \textit{Phys. Rev. Lett.} \textbf{78}, 1130 (1997)
%
\bibitem{Markovic01} N. Markovi\'{c} \textit{et al.} \textit{Phys.
Rev. B} \textbf{65}, 012501 (2001).
%
\bibitem{Daghero12} D. Daghero, \textit{et al.} \textit{Phys. Rev. Lett.} \textbf{108}, 066807 (2012).
%
\bibitem{Rowell03} J.M. Rowell, \textit{Supercond. Sci. Technol.} \textbf{16}, R17 (2003).
%
\bibitem{Yuan10} H. Yuan \emph{et al.}, J. Am. Chem. Soc.
\textbf{132}, 18402 (2010).
%
\bibitem{Nair11} Jijeesh R. Nair \emph{et al.}, \textit{Reactive and Functional Polymers} \textbf{71}, 409-416 (2011).
%
\bibitem{Inzelt10} G. Inzelt, in \emph{Electroanalytical Methods. Guide to Experiments and
Applications}, edited by F. Scholz, Springer-Verlag 2010, p. 147.
%
\bibitem{Omini} M. Omini \emph{et al.}, unpublished.
%
\bibitem{Giustino10} G. Savini, A. C. Ferrari, and F.
Giustino, \textit{Phys. Rev. Lett.} \textbf{105}, 037002 (2010).
\end{thebibliography}
\end{document}